\documentclass[conference]{IEEEtran}
\IEEEoverridecommandlockouts
\usepackage{cite}
\usepackage{amsmath,amssymb,amsfonts}
\usepackage{algorithmic}
\usepackage{graphicx}
\usepackage{textcomp}
\usepackage{xcolor}
\usepackage{subfigure}
\usepackage{stfloats}
\usepackage{amsfonts,subfigure,multicol,color,verbatim,
graphicx,cite,epsfig,amssymb,amsmath,cases,bm,algorithm,
algorithmic,xcolor,multirow,url,dsfont,bbm,tabularx}
\def\BibTeX{{\rm B\kern-.05em{\sc i\kern-.025em b}\kern-.08em
    T\kern-.1667em\lower.7ex\hbox{E}\kern-.125emX}}

\usepackage{algorithm}
\usepackage{algorithmic}
\usepackage{setspace}

\IEEEoverridecommandlockouts
\IEEEpubid{\makebox[\columnwidth]{978-1-7281-8678-8/22/\$31.00 \copyright~2022~IEEE }
	\hspace{\columnsep}\makebox[\columnwidth]{ }}

\begin{document}

\title{Low-Complexity Distributed Precoding in User-Centric Cell-Free mmWave MIMO Systems}

\author{\IEEEauthorblockN{Yingrong Zhong, Yashuai Cao and Tiejun Lv}
\IEEEauthorblockA{Key Laboratory of Trustworthy Distributed Computing and Service, Ministry of Education \\
School of Information and Communication Engineering \\
Beijing University of Posts and Telecommunications, Beijing, China\\
\{zhongyingrong, yashcao, lvtiejun\}@bupt.edu.cn}
}

\maketitle

\begin{abstract}
 User-centric (UC) based cell-free (CF) structures can provide the benefits of coverage enhancement for millimeter wave (mmWave) multiple input multiple output (MIMO) systems, which is regarded as the key technology of the reliable and high-rate services. In this paper, we propose a new beam selection scheme and precoding algorithm for the UC CF mmWave MIMO system, where a weighted sum-rate maximization problem is formulated. Since the joint design of beam selection and precoding is non-convex and tractable with high complexity, this paper designs the beam selection and precoding separately. Particularly, the proposed beam selection aims at reducing the inter-cluster inter-beam interference, then we also propose a precoding algorithm based on the weighted sum mean-square error (WSMSE) framework, where the precoding matrix can be updated in a distributed manner.  We further employ the low-rank decomposition and Neumann series expansion (NSE) to reduce the computational complexity of the precoding. Simulations and complexity analysis verify the effectiveness of the proposed algorithm with a considerable reduction in computational complexity.
\end{abstract}


\section{Introduction}
The combination of multiple input multiple output (MIMO) and millimeter-wave (mmWave) is advocated as a key solution for beyond fifth generation (B5G) to support gigabit-per-second data rates~\cite{xiao2017millimeter}. The ultra-high frequency of mmWave encounters severe path-loss and channel blockages. To offer ubiquitous network coverage, cell-free (CF) MIMO was proposed in~\cite{ngo2015cell} to improve the link reliability through aggressive deployment densification. 
Nevertheless, the limited backhaul capacity and large path-loss of remote APs pose severe challenges to the existing CF structures. As a promising remedy, a user-centric (UC) approach~\cite{buzzi2017cell} to CF systems was proposed, where each user equipment (UE) is only served by nearby access points (APs).

The unbearable power consumption, however, due to the massive radio frequency (RF) chains, hinders the development of UC CF mmWave MIMO systems. Some practical solutions, such as the hybrid precoding\cite{alonzo2019energy} and beamspace MIMO~\cite{amadori2015low}, have been proposed. In~\cite{guo2018joint}, a classic beam selection algorithm was developed for cellular beamspace MIMO systems to significantly reduce the number of RF chains meanwhile eliminating the inter-beam interference (IBI). However, in the UC CF MIMO system, in addition to the intra-cluster IBI, the inter-cluster IBI arises. On the other hand, the centralized precoding in the UC CF system causes the  considerable fronthaul signaling overhead. 

Given this backdrop, we propose in this paper a new beam selection and precoding  scheme for the UC CF mmWave MIMO system, where a weighted sum-rate maximization (WSRM) problem is formulated. This non-convex problem is quite challenging due to binary beam selection constraints and coupled variables.  Generally, the joint design of beam selection and precoding is tractable with high complexity. Therefore, we propose a practical scheme which designs beam selection and precoding separately. To be specific, we first select the beams by defining a beam energy radio, where both the intra-cluster IBI and inter-cluster IBI are considered. Then, we develop a weighted sum mean-square error (WSMSE) framework to solve the precoding design for the UC CF mmWave MIMO system, where the precoding matrix at each AP is updated in a distributed manner. Furthermore, we resort to the low-rank decomposition and Neumann series expansion (NSE) to reduce the computational complexity of the precoding optimization.

\section{System Model}\label{section:system}

\subsection{User-Centric Cell-Free Network Model}
\begin{figure}[t]
	\centering{}\includegraphics[scale=0.4]{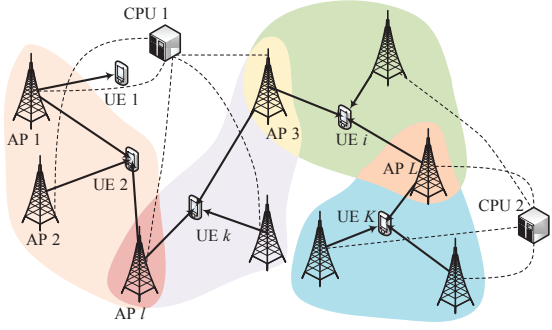}
	\caption{A UC CF network model.}
	\label{fig:signal_model}
\end{figure}

As shown in Fig. \ref{fig:signal_model}, we consider a UC CF mmWave MIMO system, where $L$ APs equipped with $N$ antennas jointly serve $K$ single-antenna UEs. According to the UC approach, each AP is connected to central processing unit (CPU) via fronthaul links and the CPUs are interconnected via the core network, the $k$-th UE is served by a subset of some nearby APs, i.e.,  $\mathcal{M}_{k}\subset \{1,...,L\}$, which is predetermined. Correspondingly, the subset of UEs served by the $l$-th AP is denoted as $\mathcal{K}_{l}\subset \{1,...,K\}$. We assume that the number of RF chains at each AP is $N_\text{{RF}}$ $(N_\text{{RF}} = K)$.
The transmit signal from the $l$-th AP is given by
\begin{align}
	\mathbf{x}_{l}=\sum_{i=1}^{K}\mathbf{D}_{il}\mathbf{w}_{il}q_{i}, \label{eq:1}
\end{align}
where $q_{i}\sim \mathcal{CN}(0,1)$ is the data symbol for the $i$-th UE, $\mathbf{w}_{il}\in \mathbb{C}^{N_{\text{RF}}\times 1}$ is the precoding vector for the $i$-th UE. The precoding matrix is $\mathbf{W}_{l}=\{\mathbf{w}_{1l}, \mathbf{w}_{2l},..., \mathbf{w}_{il}, i \subset \mathcal{K}_{l}\}$ from the $l$-th AP.  $\mathbf{D}_{il}\in \mathbb{C}^{N_{\text{RF}}\times N_{\text{RF}} }$ is used to determine whether the $l$-th AP communicates with $i$-th UE. To be specific, we define
\begin{align}
	\mathbf{D}_{il}=
	\begin{cases}
		\mathbf{I}_{N_{\text{RF}}},& l\in\mathcal{M}_{i},\\
		\mathbf{O}_{N_{\text{RF}}},& l\notin\mathcal{M}_{i}.
	\end{cases}
	\label{eq:R3}
\end{align}
The downlink received signal to the $k$-th UE is given by
\begin{align}
	y_{k}^\text{{dl}}=\sum_{l=1}^{L}
	\overline{\mathbf{h}}_{kl}^{\mathbf{H}}\sum_{i=1}^{K}\mathbf{D}_{il}\mathbf{w}_{il}q_{i}+n_k, \label{eq:R4}
\end{align}
where $n_k$ is the noise which follows the complex Gaussian distribution of $\mathcal{CN}(0,\delta^2_{\text{dl}})$. 
$\overline{\mathbf{h}}_{kl}\triangleq\mathbf{F}_{l}\mathbf{U}_{kl}\mathbf{h}_{kl}$ is the effective channel of the $l$-th AP associated with the $k$-th UE, where $\mathbf{h}_{kl}\in \mathbb{C}^{N \times 1}$ is the mmWave channel from the $k$-th UE to the $l$-th AP, $\mathbf{U}_{kl}\in \mathbb{C}^{N \times N}$ is the corresponding normalized discrete Fourier transformation (DFT) matrix
, and $\mathbf{F}_{l} \in \mathbb{C}^{N_{\text{RF}} \times N}$ is the beam selection matrix of the $l$-th AP.
By using (\ref{eq:1}), we rewrite (\ref{eq:R4}) as
\begin{align}
	\begin{aligned}
		y_{k}^\text{{dl}}&=\sum_{l=1}^{L}\overline{\mathbf{h}}_{kl}^{\mathbf{H}}\sum_{i=1}^{K}\mathbf{z}_{il}q_{i}+n_k\\
		&=\sum_{l=1}^{L}\overline{\mathbf{h}}_{kl}^{\mathbf{H}}\mathbf{z}_{kl}q_{k}+\sum_{l=1}^{L}\overline{\mathbf{h}}_{kl}^{\mathbf{H}}\sum_{i\neq k}^{K}\mathbf{z}_{il}q_{i}+n_k,
	\end{aligned}
	\label{eq:SINR}
\end{align}
where $\mathbf{z}_{il}\triangleq\mathbf{D}_{il}\mathbf{w}_{il}$ is the effective precoding vector and each AP only needs to design the precoding vectors for its served UEs.

As a result, the APs jointly serve the UEs by coherent transmission and reception~\cite{bjornson2019making}, the SINR of the $k$-th UE can be expressed as
\begin{align}
	\text{SINR}_{k}^\text{dl}= \frac { \sum_{l\in\mathcal{M}_{k}}\overline{\mathbf{h}}_{kl}^{\mathbf{H}}\mathbf{z}_{kl}\mathbf{z}_{kl}^{\mathbf{H}}\overline{\mathbf{h}}_{kl}}{\sum_{i \in\mathcal{K}_{l}\setminus\{k\}} \sum_{l\in\mathcal{M}_{i}}\overline{\mathbf{h}}_{kl}^{\mathbf{H}}\mathbf{z}_{il}\mathbf{z}_{il}^{\mathbf{H}}{\overline{\mathbf{h}}_{kl}}+\delta^2_{\text{dl}}}. \label{eq:SINR_k}
\end{align}
\subsection{Beamspace Channel Model}
The mmWave channel $\mathbf{h}_{kl}$ can be modeled as
\begin{align}
	\mathbf{h}_{kl} =
	\beta_{kl,0}\mathbf{a}(\theta_{kl,0})+\sum_{p=1}^{P}\beta_{kl,p}\mathbf{a}(\theta_{kl,p}),\label{eq:R1}
\end{align}
where the former term accounts for the line-of-sight (LoS) path, the latter accounts for $P$ non-line-of-sight (NLoS) paths, $\beta_{kl,p}$ denotes the complex gain of the $p$-th path from the $l$-th AP to $k$-th UE, $\theta_{kl,p}$ is the spatial direction from the $l$-th AP to the $k$-th UE, and $\mathbf{a}(\theta)$ is the array steering vector. By following a uniform linear array (ULA) form, $\mathbf{a}(\theta)$ is defined as
\begin{align}
	\mathbf{a}(\theta)=\frac{1}{\sqrt{N}}{[e^{-j2\pi\theta  v}]}_{v \in \mathcal{V}} , \label{eq:R2}
\end{align}
where $\mathcal{V} = \{i-\frac{N-1}{2} ; i=0,1,...,N-1\}$ is the index of antenna elements, and ${\theta}$ denotes the spatial directions.

Based on the beamspace MIMO \cite{li2017minimum}, the beamspace channel can be given by
\begin{align}
	\mathbf{U}_{kl}\mathbf{h}_{kl}=\left[\mathbf{a}(\overline{\theta}_{1}),\mathbf{a}(\overline{\theta}_{2}),...,\mathbf{a}(\overline{\theta}_{N})\right]^{\mathbf{H}}\mathbf{h}_{kl}, \label{eq:DFT}
\end{align}
where $\overline{\theta}_{n}=\frac{1}{N}(n-\frac{N+1}{2}), n\in\{1,2,...,N\}$ denotes the normalized spatial angle.

\section{Problem Formulation}
Let us concentrate on the design of the precoding and beam selection to achieve the downlink system sum-rate maximization. The downlink achievable rate of the $k$-th UE is
\begin{align}
	R_{k}=\log_{2}(1+\text{SINR}_{k}^\text{dl}). \label{eq:R_k}
\end{align}
The WSRM problem is formulated as
\begin{subequations}
	\label{eq:opt10}
	\begin{align}
		\underset{\mathbf{F}_{l},\mathbf{z}_{kl}}{\max} \quad &  \sum_{k \in \mathcal{K}_{l}}R_{k}, \label{eq:a} \\
		\text{s.t.}  \quad &  \sum_{k \in\mathcal{K}_{l}}\Vert \mathbf{z}_{kl} \Vert^2 \leq P_{\text{max}},\forall {l}, \label{eq:b}\\
		&\mathbf{F}_{l}(r,:)\mathbf{1}\leq 1,r \in \{1,2,...,N\}, \label{eq:c}\\
		&\mathbf{1}^{\mathbf{T}}\mathbf{F}_{l}(:,c)= 1,c \in \{1,2,...,N_{\text{RF}}\}, \label{eq:d}\\
		&\mathbf{F}_{l}(r,c)\in\{0,1\}, \label{eq:e}
	\end{align}
\end{subequations}
where $P_{\text{max}}$ is the maximum power of each AP; constraint (\ref{eq:c}) guarantees that each UE can only select one RF chain at most; constraint (\ref{eq:d}) indicates that each UE needs to select a single RF chain; and constraint (\ref{eq:e}) indicates the binary allocation indicator of the RF chain. These constraints ensure that the all UEs are assigned to RF chains, properly.

\section{The Proposed Algorithm}
Problem (\ref{eq:opt10})  involves the integer programming and the coupling variables, so we cannot solve it directly by using the existing convex optimization methods. To this end, we propose a practical suboptimal algorithm which designs the beam selection and precoding algorithm  separately to achieve satisfactory performance with competitive computational complexity.
\subsection{Beam Selection Scheme}
Taking into account the inter-cluster IBI presented in the UC CF system, we propose a new beam selection scheme with two basic criterion: i) to allocate the beam with the strongest gain to the UEs as much as possible, and ii) to mitigate the inter-cluster IBI. 

The proposed scheme classifies all UEs into two groups according to their strongest beams, i.e., interfering users (IUs) and non-interfering users (NIUs). Specifically, we choose the strongest beam of the $k$-th UE according to its channel gain $\{ \Vert \mathbf{h}_{1l}\Vert, \Vert \mathbf{h}_{2l} \Vert, ..., \Vert \mathbf{h}_{Kl}\Vert\},\forall l$, from the beam set of all UEs $\{b_1,b_2,...,b_{K}\}$. If the strongest beam of the $k$-th UE is different from the strongest beam of any other UEs, the $k$-th UE is selected as an NIU, otherwise it belongs to IUs. Our proposed beam selection scheme for different categories of UEs is given by:
\begin{enumerate}
	\item For each AP, if the UE belongs to the NIU, it will be assigned to the strongest beam to achieve maximum sum-rate, otherwise we select the most suitable beam from remaining beams to reduce the intra-cluster IBI, as the interference-aware beam selection (IA-BS) scheme proposed in~\cite{7438800}.
	\item When we consider the IBI caused by other UEs in other different clusters based on 1), we use $E_{k,n}^{\mathrm{in}}$ and $E_{k^{\prime},n}^{\mathrm{out}}$ to represent the intra-cluster energy (ICE) of the $k$-th UE and the out-of-cluster energy (OCE) of the $k^{\prime}$-th UE corresponding to the overlapped beam  respectively, where the $k^{\prime}$-th UE is not in the cluster of the $k$-th UE. Define the ratio of the ICE versus the OCE as $r_{n}=\frac{E^{\mathrm{out}}_{k^{\prime},n}}{E^{\mathrm{in}}_{k,n}}$. If $r_{n}$ is greater than a predetermined threshold $\gamma_{\mathrm{th}}$, we select the most suitable beam for the $k$-th UE from the remaining beams one by one until the suboptimal beam is selected. Otherwise, we regard the inter-cluster IBI as trivial and thus can be ignored.
\end{enumerate}
\subsection{Design of Precoding Matrix}
For the conventional centralized precoding, the CPU can design the precoding with the global channel state information (CSI) so that the APs can cancel out the multi-user interference. However, this operation is not applicable to the downlink CF MIMO system where each AP only knows local CSI knowledge.
To derive a practical algorithm, we propose a WSMSE framework to facilitate distributed implementation of precoding based on the limited CSI sharing. Differently from traditional WSMSE methods, our proposed WSMSE framework avoids the computational intensive matrix inversion by applying the low-rank decomposition and NSE.

Given the beam selection results, the WSRM problem reduces to
\begin{align}
	\underset{\mathbf{z}_{kl}}{\max} \quad &  \sum_{k \in \mathcal{K}_{l}}R_{k}, \label{eq:opt10_2} \\
	\text{s.t.}  \quad &  \nonumber (\ref{eq:b}).
\end{align}

We first transform problem (\ref{eq:opt10_2}) into a WSMSE problem:
\begin{align}
	\underset{\mathbf{z}_{kl},\mu_k,\alpha_k}{\min} \quad &  \sum_{k \in \mathcal{K}_{l}}\alpha_{k}E_{k}(\mu_k)-\log{\alpha_{k}}, \label{eq:opt11} \\
	\text{s.t.}  \quad &  \nonumber (\ref{eq:b}),
\end{align}
where $\alpha_{k}> 0,k\in\mathcal{K}_{l}$,  $E_{k}(\mu_k)\triangleq \mathbb{E}\{|{\mu_{k}}y_{k}^\text{{dl}}-q_{k}|^2\}$ is the mean square error (MSE) of the $k$-th UE, $\mu_{k}$ is the receiver coefficients. Next, we solve problem (\ref{eq:opt11}) by the alternating iterative methods.

Given $\alpha_{k}$ and $\mathbf{z}_{kl}$, we optimize $\mu_{k}$ by the first-order derivative of the MSE:
\begin{align}
	\mu_k^{\mathrm{opt}}=\frac{\sum_{l\in\mathcal{M}_{k}}\overline{\mathbf{h}}_{kl}^{\mathbf{H}}\mathbf{z}_{kl}}
	{\sum_{i \in \mathcal{K}_{l}}\sum_{l\in\mathcal{M}_{i}}\overline{\mathbf{h}}_{kl}^{\mathbf{H}}\mathbf{z}_{il}\mathbf{z}_{il}^{\mathbf{H}}
		\overline{\mathbf{h}}_{kl}+\delta_{\text{dl}}^2}. \label{eq:opt12}
\end{align}
Substituting (\ref{eq:opt12}) into $E_{k}(\mu_{k})$ yields
\begin{align}
	E_{k}(\mu_k^{\mathrm{opt}})=\frac{\sum_{i \in\mathcal{K}_{l}\setminus\{k\}}\sum_{l\in\mathcal{M}_{i}}\overline{\mathbf{h}}_{kl}^{\mathbf{H}}\mathbf{z}_{il}\mathbf{z}_{il}^{\mathbf{H}}\overline{\mathbf{h}}_{kl}+\delta_{\text{dl}}^2}
	{\sum_{i \in \mathcal{K}_{l}}\sum_{l\in\mathcal{M}_{i}}\overline{\mathbf{h}}_{kl}^{\mathbf{H}}\mathbf{z}_{il}\mathbf{z}_{il}^{\mathbf{H}}\overline{\mathbf{h}}_{kl}+\delta_{\text{dl}}^2}. \label{eq:opt13}
\end{align}
Given $\mathbf{z}_{kl}$ and $\mu_k$, we set the first-order derivative of the problem (\ref{eq:opt11}) to zero with respect to $\alpha_{k}$. Then $\alpha_{k}$ is updated as
\begin{align}
	\alpha_{k}^{\mathrm{opt}}=1+\frac{\sum_{l\in\mathcal{M}_{k}}\overline{\mathbf{h}}_{kl}^{\mathbf{H}}\mathbf{z}_{kl}\mathbf{z}_{kl}^{\mathbf{H}}\overline{\mathbf{h}}_{kl}}{\sum_{i \in\mathcal{K}_{l}\setminus\{k\}}\sum_{l\in\mathcal{M}_{i}}\overline{\mathbf{h}}_{kl}^{\mathbf{H}}\mathbf{z}_{il}\mathbf{z}_{il}^{\mathbf{H}}\overline{\mathbf{h}}_{kl}+\delta_{\text{dl}}^2}. \label{eq:opt14}
\end{align}
Eqns. (\ref{eq:opt12}) and (\ref{eq:opt14}) indicate that each UE needs to collect its associated local CSI and effective precoding vector $\mathbf{z}_{kl}$, which are used to update $\alpha_{k}^{\mathrm{opt}}$ and $\mu_{k}^{\mathrm{opt}}$ at the UEs. Given $\mu_{k}$ and $\alpha_{k}$, problem (\ref{eq:opt10}) can be rewritten as
\begin{align}
	\underset{\mathbf{z}_{kl}}\min& \quad  {F(\mathbf{z}_{kl})}\triangleq {\sum_{k \in \mathcal{K}_{l}}\alpha_{k}E_{k}(\mu_k)}, \label{eq:opt15}\\
	\text{s.t}.& \quad   \nonumber (\ref{eq:b}),
\end{align}
where $F(\mathbf{z}_{kl})$ is
\begin{align}
	\begin{aligned}
		F(\mathbf{z}_{kl})=& {\sum_{k \in \mathcal{K}_{l}}}\alpha_{k}\bigg[ \mu_{k}^{*}\Big(\sum_{i \in \mathcal{K}_{l}}\sum_{l\in\mathcal{M}_{i}}\overline{\mathbf{h}}_{kl}^{\mathbf{H}}\mathbf{z}_{il}{\mathbf{z}_{il}}^{\mathbf{H}} \overline{\mathbf{h}}_{kl}+\delta_{\text{dl}}^2\Big) \\ &\mu_{k}-\mu_{k}^{*}\sum_{l\in\mathcal{M}_{k}}\overline{\mathbf{h}}_{kl}^{\mathbf{H}}\mathbf{z}_{kl}-{\mathbf{z}_{kl}}^{\mathbf{H}}
		\overline{\mathbf{h}}_{kl}\mu_{k}+1\bigg].
	\end{aligned}
	\label{eq:opt16}
\end{align}
By introducing the Lagrange multiplier $\lambda$ for the constraint (\ref{eq:b}), the Lagrangian function of problem (\ref{eq:opt15}) is given by
\begin{align}
	\mathcal{L}(\mathbf{z}_{kl},\lambda)\triangleq F(\mathbf{z}_{kl})+\lambda\left(\sum_{k \in \mathcal{K}_{l}}\Vert \mathbf{z}_{kl}\Vert^2-P_{\text{max}}\right)
	. \label{eq:opt17}
\end{align}
By setting $\frac{\partial{\mathcal{L}(\mathbf{z}_{kl},\lambda)}}{\partial{\mathbf{z}_{kl}}}=0$ and $\frac{\partial{\mathcal{L}(\mathbf{z}_{kl},\lambda)}}{\partial{\lambda}}=0$, we have
\begin{align}
	\begin{aligned}
		\mathbf{z}_{kl}^{\mathrm{opt}}=\left( \sum_{k \in \mathcal{K}_{l}}\alpha_{k}|\mu_{k}|^2 \overline{\mathbf{h}}_{kl}\overline{\mathbf{h}}_{kl}^{\mathbf{H}}+\lambda\mathbf{I}_{N_\text{RF}}\right)^{-1} \times {
			\alpha_{k}{\mu}_{k}}\overline{\mathbf{h}}_{kl},
	\end{aligned}
	\label{eq:opt18}
\end{align}
where $\mathbf{I}_{N_\text{RF}} $ is an ${N_\text{RF}}\times {N_\text{RF}}$ identity matrix. The update of $\mathbf{z}_{kl}^{\mathrm{opt}}$ needs to require $\alpha_{k}^{\mathrm{opt}}$ and $\mu_{k}^{\mathrm{opt}}$, which are collected by the $l$-th AP via the feedback channel between the UEs.

Now we proceed to find $\lambda$ to solve problem (\ref{eq:opt15}).
For notational brevity, we define $\mathbf{H}=\sum_{k \in \mathcal{K}_{l}}\alpha_{k}|\mu_{k}|^2 \overline{\mathbf{h}}_{kl}\overline{\mathbf{h}}_{kl}^{\mathbf{H}}$, $\mathbf{H}$ is a positive semi-definite matrix, and its rank is $\mathrm{rank}(\mathbf{H})={N_\text{RF}}^{\prime}\leq {N_\text{RF}}$. To reduce the computational complexity, we apply the eigenvalue decomposition to $\mathbf{H}$ yields
\begin{align}
	\mathbf{H} = \left[\mathbf{J}_{1},\ \mathbf{J}_{2}\right]\begin{bmatrix}\mathbf{\Sigma}_{1}&\ \\ \ &\mathbf{\Sigma}_{2}\end{bmatrix} \left[\mathbf{J}_{1},\ \mathbf{J}_{2}\right]^{\mathbf{H}}, \label{eq:opt20}
\end{align}
where $\mathbf{J}_{1}$ is the first ${N_\text{RF}}^{\prime}$ singular vectors; its corresponding positive eigenvalue is $\mathrm{Diag}(\epsilon_1,\epsilon_2,...,\epsilon_{{N_\text{RF}}^{\prime}})$
with $\epsilon_{1}\ge\epsilon_{i}\ge\epsilon_{{N_\text{RF}}^{\prime}}$; $\epsilon_{i}$ is the $i$-th diagonal element in $\mathbf{\Sigma}_{1}$, $\epsilon_{1}$ and $\epsilon_{{N_\text{RF}}^{\prime}}$ are the maximum eigenvalue and minimum eigenvalue of $\tilde{\mathbf{H}}$; $\mathbf{J}_{2}$ is the remaining $({N_\text{RF}}-{N_\text{RF}}^{\prime})$ singular vectors corresponding to the eigenvalue matrix $\mathbf{\Sigma}_{2}$. We thus replace (\ref{eq:opt20}) by
\begin{align}
	\tilde{\mathbf{H}} = \mathbf{J}_{1}\mathbf{\Sigma}_{1}\mathbf{J}_{1}^{\mathbf{H}}. \label{eq:opt21}
\end{align}
Therefore, Eq. (\ref{eq:opt18}) can be rewritten as
\begin{align}
	\mathbf{z}_{kl}^{\mathrm{opt}}=(\tilde{\mathbf{H}}+\lambda\mathbf{I}_{N_\text{RF}})^{-1}{
		\alpha_{k}{\mu}_{k}}\overline{\mathbf{h}}_{kl}. \label{eq:z}
\end{align}

Based on the slackness condition for power constraints and the Bisection method~\cite{9105111}, we search the optimal $\lambda$ in range $[\lambda_{\min}, \lambda_{\max}]$ and substituting it into $\mathbf{z}_{kl}$ to update the precoding vector.

Note that the $\mathbf{z}_{kl}^{\mathrm{opt}}$ can be updated in a distributed manner.
Specifically, each AP first feeds back the local CSI to its served UEs. Then, the UEs update $\{\mu_{k},\forall k\}$ and $\{\alpha_{k},\forall k\}$ by collecting the knowledge of local CSI and $\{\mathbf{z}_{kl}, \forall k,l\}$.  Finally, the UEs feed these updated parameters back to the APs. Each AP can update $\{\mathbf{z}_{kl}, \forall k,l\}$ by (\ref{eq:opt18}) without exchanging the local CSI with other APs. 
Therefore, the distributed WSMSE scheme reduces the overhead of feedback signaling.

\begin{algorithm}[t]
	\small
	\caption{The proposed beam selection and precoding algorithm.}
	\label{alg:A1}
	\begin{algorithmic}[1]
		\REQUIRE 
		\STATE Initialize  $\{\mathbf{z}_{kl}, \forall {k,l}\}$ such that $\sum_{k \in \mathcal{K}_{l}}\Vert \mathbf{z}_{kl} \Vert^2 \leq P_{\text{max}}$ and iteration indexes $i, j=0$.
		\REPEAT
		\STATE {Set $i \leftarrow i+1$;}
		\STATE {Select the suitable beams $\{b_{k}^{(i)}, \forall k\}$ for all UEs by the IA-BS scheme;}
		\STATE {Select the suboptimal beams and release the overlapped beams if $r_{n}^{(i)} > \gamma_{\mathrm{th}}$;}
		\UNTIL{$r_{n}^{(i)} -\gamma_{\mathrm{th}} \leq 0$}
		\ENSURE
		\REPEAT
		\STATE {Set $j \leftarrow j+1$;}
		\STATE {Update $\{\mu_k^{(j)}, \forall {k}\}$ by (\ref{eq:opt12});}
		\STATE {Update $\{\alpha_k^{(j)}, \forall {k}\}$ by (\ref{eq:opt14});}
		\STATE {Update $\beta^{(j)}$ by (\ref{eq:opt29}) and $\lambda^{(j)}$ through the Bisection method;}
		\STATE {Update $\{\mathbf{z}_{kl}^{(j)}, \forall {k,l}\}$ by (\ref{eq:opt28});}		
		\UNTIL {Problem (\ref{eq:opt10}) converges.}
	\end{algorithmic}
\end{algorithm}
\subsection{Low-Complexity Precoding Design}
To reduce the computational complexity of the matrix inversion in Eq. (\ref{eq:z}), we resort to the NSE method according to \cite{minango2017low}. To be specific, we rewrite (\ref{eq:z}) as
\begin{align}
	\mathbf{z}_{kl}^{\mathrm{opt}}={\alpha_{k}{\mu}_{k}{\mathbf{Z}}^{-1}\overline{\mathbf{h}}_{kl}}, \label{eq:opt26}
\end{align}
where
$\mathbf{Z}\triangleq\tilde{\mathbf{H}}+\lambda\mathbf{I}_{N_\text{RF}}$.

The NSE of $\mathbf{Z}^{-1}$ is defined as
\begin{align}
	\mathbf{Z}^{-1}\approx\beta\sum_{t=0}^{\infty}\Big(\mathbf{I}_{N_\text{RF}}-\beta \mathbf{Z}\Big)^{t}, \label{eq:opt27}
\end{align}
where $\beta$ is a scaling factor. When $\lim_{t\to\infty}(\mathbf{I}_{N_\text{RF}}-\beta \mathbf{Z})^{t}\to \mathbf{O}$, the above approximate expansion will work. Hence, the precoding vector $\mathbf{z}_{kl}^{\mathrm{opt}}$ can be expressed as

\begin{align}
	\begin{aligned}
		\mathbf{z}_{kl}^{\mathrm{opt}(t)}&\approx{\alpha_{k}{\mu}_{k}\beta}\Big(\mathbf{I}_{N_\text{RF}}+(\mathbf{I}_{N_\text{RF}}-\beta \mathbf{Z})+(\mathbf{I}_{N_\text{RF}}-\\ &\beta \mathbf{Z})^2+... +(\mathbf{I}_{N_\text{RF}}-\beta \mathbf{Z})^{t}\Big)\overline{\mathbf{h}}_{kl},
	\end{aligned}
	\label{eq:opt28}
\end{align}
where $\beta$ can be taken as $\frac{2}{\kappa_{\max}(\mathbf{Z})+\kappa_{\min}(\mathbf{Z})}$ \cite{zhu2015matrix} with $\kappa_{\max}(\mathbf{Z})$ and $\kappa_{\min}(\mathbf{Z})$ being the largest and the smallest eigenvalues of $\mathbf{Z}$. According to the eigenvalues in (\ref{eq:opt20}), the $\kappa_{\max}(\mathbf{Z})$ and $\kappa_{\min}(\mathbf{Z})$ can be calculated by
\begin{align}
	\kappa_{\min}(\mathbf{Z})=\epsilon_{{N_\text{RF}}^{\prime}}+\lambda,\quad
	\kappa_{\max}(\mathbf{Z})=\epsilon_{1}+\lambda.  \label{eq:opt29}
\end{align}

The proposed beam selection and precoding algorithm is provided in Algorithm \ref{alg:A1}.
\subsection{Computational Complexity}
We provide the complexity analysis: 1) The complexity of the beam selection algorithm is $\mathcal{O}(KNN_{\text{RF}})$; 2) The calculation of $\mu_k^{\mathrm{opt}}$ and $\alpha_k^{\mathrm{opt}}$ needs $\mathcal{O}(K^2)$ and $\mathcal{O}(K)$, respectively; 3) We need $\mathcal{O}({N_{\text{RF}}}^3+ K^3{N_{\text{RF}}}^{\prime} + \log{\frac{\lambda_{\max}}{T_\text{tol}}})$ multiplications to decompose $\mathbf{H}$, and find $\lambda$ in the Bisection method, where $T_\text{tol}$ denotes the searching tolerance; 4) Since the complexity of the matrix inversion is the same level as the matrix multiplication, the complexity of $\mathbf{Z}^{-1}$ is $\mathcal{O}({N_{\text{RF}}}^3)$.
When $N \to \infty$, the total asymptotic complexity of our proposed algorithm is $\mathcal{O}(I_\text{iter}L(KNN_{\text{RF}}))$~\cite{guo2018joint}, where $I_\text{iter}$ is the number of convergence iterations.
Compared with the centralized CF scheme with the ZF precoding, the computational complexity of the proposed beam selection and precoding algorithm is reduced from $\mathcal{O}(I_\text{iter}L(N^2KN_{\text{RF}}))$ to $\mathcal{O}(I_\text{iter}L(KNN_{\text{RF}}))$.  In the proposed method, the matrix inversion is transformed into matrix multiplications which are suitable for hardware calculation.

\section{Numerical Results}\label{section:sim}
	%
	%
\begin{figure}[t]
	\centering
	\includegraphics[scale=0.5]{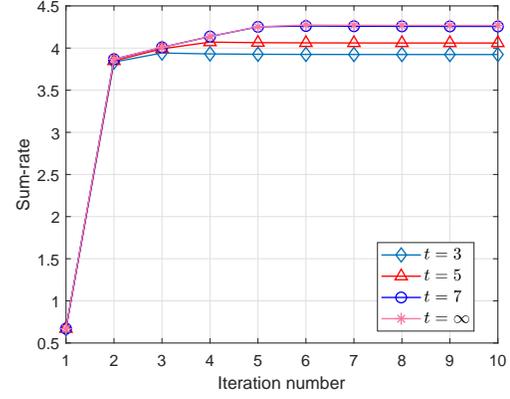}
	\caption{Convergence behavior of the proposed algorithm.}
	\label{fig:sim1}
\end{figure}

\begin{figure}[t]
	\centering
	\includegraphics[scale=0.5]{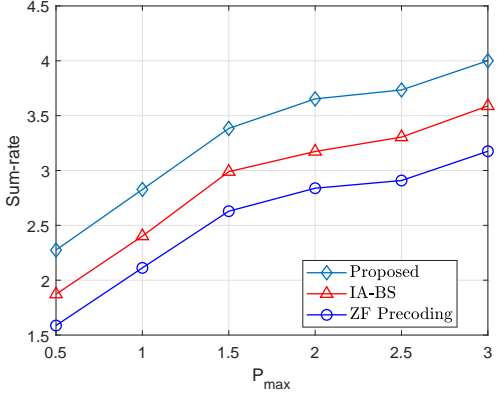}
	\caption{Sum-rate versus $P_{\max}$ with $N=32$ and $K=8$.}
	\label{fig:sim2}
\end{figure}

\begin{figure}[t]
	\centering
	\includegraphics[scale=0.5]{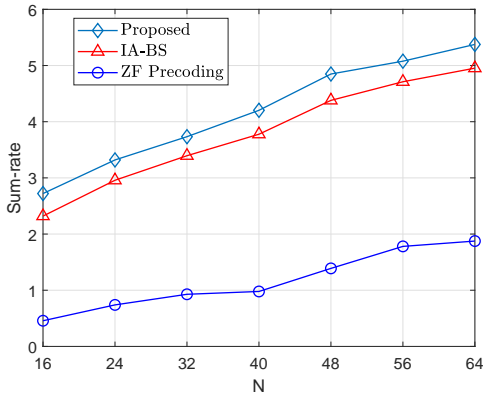}
	\caption{Sum-rate versus $N$ with $P_{\max}=1$ W and $K=8$.}
	\label{fig:sim3}
\end{figure}

In this section, we show the effectiveness of the proposed method through the system sum-rate. We consider a square area $250\times250$ $\text{m}^2$, where $K = 8$ UEs and $L = 32$ APs are randomly located. According to \cite{alonzo2017cell} and \cite{haneda20165g}, the channel gain is taken as $\alpha_{k,l}\sqrt {PL(d_{k,l})}$, where $\alpha_{k,l} \sim \mathcal{CN}(0,1)$, and $PL(d) = -20\log_{10}{(\frac{4\pi f}{c})}-10n(1+\frac{bcf}{f_0})\log_{10}(d)-X_{\delta}$ is path-loss model, with $X_{\delta} \sim \mathcal{CN}(0,4.2)$. More details about the path-loss parameter settings can be found in \cite{haneda20165g}, such as UMi (Urban Microcellular). We set $\delta^2_{\text{dl}} = -85$ dBm, $N = 16$, $N_{\text{RF}} = 8$, and $P_\text{max}=1$ W.

Fig. \ref{fig:sim1} plots the convergence of our proposed algorithm versus different NSE orders. It is seen that our proposed algorithm can converge after about $5$ iterations. As the number of NSE orders increases to $t=7$, the sum-rate after convergence is nearly close to that achieved by $t=\infty$. 

Fig. \ref{fig:sim2} shows the impact of different AP transmit powers on the sum-rate achieved by our proposed algorithm. In addition to the proposed beam selection algorithm and precoding scheme, two benchmark schemes are evaluated for comparison. The first benchmark employs the beam selection of the IA-BS scheme, and then updates the precoding matrix by using our proposed precoding scheme. The second benchmark is performed by using the ZF precoding scheme with our proposed beam selection algorithm. From the results, we see that as $P_\text{max}$ increases from $0.5$ W to $3$ W, the system sum-rate also increases. The best performance is achieved by the proposed algorithm, since it considers the inter-cluster IBI, while other schemes fail to reduce the inter-cluster IBI. This demonstrates the effectiveness of our proposed algorithm.

Fig. \ref{fig:sim3} evaluates the impact of our proposed algorithm versus different numbers of AP antennas. Compared with the IA-BS and ZF precoding schemes, we see that the system sum-rate increases as $N=16$ increases to $N=64$. Our proposed algorithm greatly outperforms the ZF precoding scheme, which reveals a significant precoding gain of our proposed scheme. This demonstrates again the validity of our proposed algorithm.

\section{Conclusion}\label{section:con}
In this paper, a low-complexity beam selection and precoding scheme algorithm was proposed to maximize the UC CF mmWave MIMO system sum-rate. Specifically, we proposed a beam selection to reduce the inter-cluster IBI and devised a WSMSE framework to solve precoding design problem. To reduce the computational complexity of the precoding, we also employed the low-rank decomposition and NSE for the WSMSE framework. Simulation results have demonstrated that the convergence and effectiveness of the proposed algorithm.



\bibliographystyle{IEEEtran}
\bibliography{ciations}

\begin{thebibliography}{10}
\providecommand{\url}[1]{#1}
\csname url@samestyle\endcsname
\providecommand{\newblock}{\relax}
\providecommand{\bibinfo}[2]{#2}
\providecommand{\BIBentrySTDinterwordspacing}{\spaceskip=0pt\relax}
\providecommand{\BIBentryALTinterwordstretchfactor}{4}
\providecommand{\BIBentryALTinterwordspacing}{\spaceskip=\fontdimen2\font plus
\BIBentryALTinterwordstretchfactor\fontdimen3\font minus
  \fontdimen4\font\relax}
\providecommand{\BIBforeignlanguage}[2]{{%
\expandafter\ifx\csname l@#1\endcsname\relax
\typeout{** WARNING: IEEEtran.bst: No hyphenation pattern has been}%
\typeout{** loaded for the language `#1'. Using the pattern for}%
\typeout{** the default language instead.}%
\else
\language=\csname l@#1\endcsname
\fi
#2}}
\providecommand{\BIBdecl}{\relax}
\BIBdecl

\bibitem{xiao2017millimeter}
M.~Xiao \emph{et~al.}, ``Millimeter wave communications for future mobile
  networks,'' \emph{IEEE J. Sel. Areas Commun.}, vol.~35, no.~9, pp.
  1909--1935, Jun. 2017.

\bibitem{ngo2015cell}
H.~Q. Ngo \emph{et~al.}, ``Cell-free massive {MIMO}: {Uniformly} great service
  for everyone,'' in \emph{Proc. IEEE SPAWC}, Jun. 2015, pp. 201--205.

\bibitem{buzzi2017cell}
S.~Buzzi \emph{et~al.}, ``Cell-free massive {MIMO}: {User-centric} approach,''
  \emph{IEEE Wirel. Commun. Lett.}, vol.~6, no.~6, pp. 706--709, Aug. 2017.

\bibitem{alonzo2019energy}
M.~Alonzo \emph{et~al.}, ``Energy-efficient power control in cell-free and
  user-centric massive {MIMO} at millimeter wave,'' \emph{IEEE Trans. Green
  Commun. Netw.}, vol.~3, no.~3, pp. 651--663, Mar. 2019.

\bibitem{amadori2015low}
P.~V. Amadori \emph{et~al.}, ``Low {RF}-complexity millimeter-wave
  beamspace-{MIMO} systems by beam selection,'' \emph{IEEE Trans. Commun.},
  vol.~63, no.~6, pp. 2212--2223, May. 2015.

\bibitem{guo2018joint}
R.~Guo \emph{et~al.}, ``Joint design of beam selection and precoding matrices
  for {mmWave} {MU-MIMO} systems relying on lens antenna arrays,'' \emph{IEEE
  J. Sel. Top. Signal Process.}, vol.~12, no.~2, pp. 313--325, Apr. 2018.

\bibitem{bjornson2019making}
E.~Bj{\"o}rnson \emph{et~al.}, ``Making cell-free massive {MIMO} competitive
  with {MMSE} processing and centralized implementation,'' \emph{IEEE Trans. on
  Wirel. Commun.}, vol.~19, no.~1, pp. 77--90, Jan. 2020.

\bibitem{li2017minimum}
D.~Li \emph{et~al.}, ``Minimum interference beam selection for millimeter wave
  beamspace {MIMO} system,'' in \emph{Proc. Int. Conf. Comm. Netw.}, Jun. 2017,
  pp. 141--152.

\bibitem{7438800}
X.~Gao \emph{et~al.}, ``Near-optimal beam selection for beamspace {mmWave}
  {Massive} {MIMO} {systems},'' \emph{IEEE Commun. Lett.}, vol.~20, no.~5, pp.
  1054--1057, Mar. 2016.

\bibitem{9105111}
Z.~Li \emph{et~al.}, ``Weighted sum-rate maximization for multi-irs aided
  cooperative transmission,'' \emph{IEEE Wirel. Commun. Lett.}, vol.~9, no.~10,
  pp. 1620--1624, Jun. 2020.

\bibitem{minango2017low}
J.~Minango \emph{et~al.}, ``Low-complexity {MMSE} detector based on the
  first-order neumann series expansion for massive {MIMO} systems,'' in
  \emph{Proc. IEEE Latin-Amer. Conf. Commun. (LATINCOM)}, Dec. 2017.

\bibitem{zhu2015matrix}
D.~Zhu \emph{et~al.}, ``On the matrix inversion approximation based on
  {Neumann} series in massive {MIMO} systems,'' in \emph{Proc. IEEE Int. Conf.
  Commun. (ICC)}, Jun. 2015, pp. 1763--1769.

\bibitem{alonzo2017cell}
M.~Alonzo \emph{et~al.}, ``Cell-free and user-centric massive {MIMO} at
  millimeter wave frequencies,'' in \emph{Proc. IEEE Annu. Int. Symp. Pers.,
  Indoor, Mobile Radio Commun. (PIMRC)}, Oct. 2017.

\bibitem{haneda20165g}
K.~Haneda \emph{et~al.}, ``{5G} {3GPP}-like channel models for outdoor urban
  microcellular and macrocellular environments,'' in \emph{Proc. IEEE Veh.
  Technol. Conf. (VTC spring)}, May. 2016.

\end{thebibliography}

\end{document}